\documentclass[10pt,english]{article}
\usepackage[margin=1in]{geometry}
\usepackage{subfigure}
\usepackage{graphicx}
\usepackage{float}
\usepackage{amsmath}
\usepackage{dsfont}
\usepackage{setspace}
\usepackage{cite}
\usepackage{hyperref}
\usepackage{authblk}
\usepackage{orcidlink}

\usepackage{listings}
\usepackage{babel}

\title{O-SCOPE: a Raspberry Pi Pico oscilloscope}

\author{Gabriel Andrecut\orcidlink{0009-0003-6729-4138} and Mircea Andrecut\orcidlink{0000-0002-4172-7295}\thanks{mircea.andrecut@gmail.com}}

\date{September 1, 2026}

\begin{document}
\maketitle

\bigskip
\bigskip

\begin{abstract}

In this paper we provide an introduction to some of the efficient features of the Raspberry Pi Pico (RP2) family of microconroller boards (RP2040 and RP2350). 
More exactly, we show how to use PIO (Programmable Input/Output) and DMA (Direct Memory Access) to program the RP2 boards 
for efficient DAC (Digital to Analog) and ADC (Analog to Digital) conversions. Using OP-AMPS (Operational Amplifiers) we also show how to build a minimal electronic frontend interface  
for signal input/output. Finally, we integrate all these components into a product, the O-SCOPE, which is a 
practical and easy to use oscilloscope and wave function generator. The RP2 component of the O-SCOPE is built using Micropython, 
while the user interface is developed using vanilla Javascript and it runs on the web browser of any computer, interacting with the RP2 board via the Web Serial API. 
The O-SCOPE web interface is hosted on GitHub Pages such that it is directly accessible everywhere using a web browser, such that no local installation is required. 
Also, the RP2 software component is freely distributed on GitHub using the UF2 (USB Flashing Format) mass storage file, and it can be installed by simply dragging the file on the RP2 boards. 
These features make the O-SCOPE easily accessible, transforming it into a very low cost and useful tool for laboratories everywhere in the world. 

\bigskip

\end{abstract}

\bigskip
\bigskip

\section{Introduction}

Microcontrollers are single silicon chip computers typically used to control sensors, circuits and peripheral devices, and they are key components in the development of the Internet of Things (IOT). 
The Raspberry Pi Pico (RP2) is a low-cost, high-performance microcontroller board using the ARM Cortex-M0+ (RP2040) \cite{key-1} or the ARM Cortex-M33 (RP2350) \cite{key-2} chips produced by the Raspberry Pi Foundation. 
The RP2350 version delivers a significantly increased performance over the original RP2040 board, thanks to its newer design. 
A direct comparison between the two boards is shown in Table 1. 
The hardware Floating Point Unit (FPU) of the Cortex-M33, and the increased RAM, can significantly speed up some signal processing and mathematical heavy applications. 

\begin{table}[t]
\centering
\caption{A comparison between the current RP2 boards.}
\medskip
\begin{tabular}{l|c|c}
\hline
Feature & Raspberry Pi Pico (RP2040) & Raspberry Pi Pico 2 (RP2350) \\
\hline
Main CPU & Dual-core Arm Cortex-M0+ & Dual-core Arm Cortex-M33  \\
Alternative CPU & None & Dual-core Hazard3 Risc-V  \\
Clock speed & 133 MHz & 150 MHz \\
Floating point & Software-implemented & Dedicated hardware FPU \\
On chip SRAM & 256 KB & 520 KB \\
GPIO pins & 30 & 30 \\
Security features & None & ARM TrustZone
\end{tabular}
\end{table}

Here we explore the use of PIO (Programmable Input/Output) and DMA (Direct Memory Access) for programming the RP2 boards, in order to perform 
efficient DAC (Digital to Analog) and ADC (Analog to Digital) conversions \cite{key-3}. 
Using OP-AMPS (operational amplifiers) \cite{key-4}, we also develop a minimal electronic frontend interface for signal input/output, 
and we integrate all these components into a product, the O-SCOPE, which is a practical and easy to use two channel oscilloscope, and two channel wave function generator (WFG). 
The O-SCOPE software is built using Micropython \cite{key-5} on the RP2 side, while the user 
interface is developed in vanilla Javascript, and it runs on the web browser of any computer, interacting with the RP2 via the Web Serial API \cite{key-6}. 
The O-SCOPE web interface is hosted on GitHub Pages such that it is directly accessible everywhere, and no local installation is required. 
The RP2 software component is also distributed on GitHub using a UF2 (USB Flashing Format) mass storage file, which can be installed by simply dragging it on the RP2 boards. 
These features make the O-SCOPE accessible everywhere in the world, transforming it into an useful tool for physics and engineering student laboratories. 

In this paper we will mostly refer to the newer RP2350 board, but we will explicitly specify the differences from RP2040 when they appear. 
The O-SCOPE software exploits the PIO and DMA capabilities of the RP2 boards, and as a consequence it 
runs equally efficient on both boards, making the older RP2 board a cheaper alternative for this project.

\section{The hardware environment}

The RP2 board is an impressive piece of hardware, featuring 30 General-Purpose Input/Output (GPIO) pins, and other 10 additional pins for: power supply, ADC voltage reference, and analog/digital ground 
\cite{key-7}, \cite{key-8}. 
The GPIO pins 26, 27 and 28 correspond to the integrated 12-bit ADC inputs (GPIO 29 is also an ADC input, but it is internally used only).  
Here we will use GPIO 26 and 27 to build two input channels for the O-SCOPE, however the same procedure can be used to extend the O-SCOPE to three input channels if necessary. 
Unfortunately, the RP2 board does not provide an output DAC interface, but we will build such an interface ourselves using the GPIO pins from 0 to 15, in order to define two WFG channels. 
Additionally, we will use the GPIO 16 and GPIO 18 for defining two PWM (pulse width modulation) channels which can be also useful for different applications. 
The board is powered via the micro-USB connector, and 
pin 36 will be used as the $3.3$V power supply for the input/output interfaces. The analog ground is provided by pin 33, and the digital ground by any of the pins: 3, 8, 13, 18, 23, 28, 38. 
The micro-USB port is used for supplying power and for communicating with the host computer. On the board there is also a button (BOOTSEL, which is short for "boot selection") that must be 
used for installing and updating the firmware. We strongly recommend to read the official documentation before starting to build the project, such that potential errors will be avoided.  

\section{The software environment}

Microcontrollers do not have an operating system, and they can be programmed to run only a single task. RP2 can be programmed both in C/C++ \cite{key-9} and Micropython \cite{key-5}. 
As mentioned in the introduction, we will use Micropython, and in this case Thonny is a very useful Integrated Development Environment (IDE) 
for establishing a communication between the host computer and the RP2 board during the development \cite{key-10}. 
Micropython is an implementation of Python 3 which includes a subset of the Python standard library. 
While Micropython has a small footprint, it is still reasonably full-featured and supports most of Python's syntax, but it also adds many features specific to microcontrollers. 
Micropython must be installed on the RP2 before the board can be used, and the installation can be done from the Thonny IDE. 
For packaging the software into a UF2 (USB Flashing Format) file we will use the Micropython cross compiler utility (mpy-cross) to pre-compile the python files into bytecode suitable for mass distribution. 

The Web Serial API \cite{key-6} is used to establish the communication between the RP2 and the browser running on the host computer. 
Here we also build a simple user interface for controlling the oscilloscope and to visualize the signal and the measurements. 
This section only requires a Web Serial capable browser, such as Chrome, while the rest of the software is written in vanilla Javascript, HTML, and Canvas for high 
performance visualization.

\section{Preliminaries}

In this section we briefly discuss some of the GPIO, ADC, PWM, PIO and DMA aspects of the RP2 microcontroller boards, with the purpose of 
emphasizing the main features that will be used in our project development. Since the amount of related information is quite complex, 
the interested reader is invited to consult the official documentation for more details \cite{key-7}, \cite{key-8}. 
In order to help the reader, here we will only provide a short guiding summary.

In Micropython, a Pin object is used to control GPIO pins, and it has methods to set the mode of the pin as IN or OUT, and to get and set the digital logic level.
The RP2 has 29 multifunction GPIO pins connectable to external devices, including three software selectable ADC pins (GPIO 26, 27, 28). 

When working with GPIO pins, we can write binary digital data, and turn a pin ON, which provides $3.3$V available on the pin we specify. 
Alternatively, we can turn the pin OFF, which provides 0V on the pin. By using the GPIO pins we can also read digital data 
corresponding to different forms of information, similarly represented as either $3.3$V (True) or 0V (False). Thus, the GPIO pins can be used to connect
the board to other devices, or to control other electronic components.

When we set up a GPIO pin as an input, the GPIO has to distinguish between three states: 
low state (0V); high state ($3.3$V); and floating state (uncertain value).
The initial configuration of the GPIO pins is the floating state, since the pins are disconnected, and as a consequence their voltage is not a steady state. 
Therefore, in order to set the GPIO pins state to a reliable reference voltage we need to pull the voltage either up or down. 
RP2 has internal resistors that can be used to pull up or down the pins as required by the implementation logic.

Micropython also provides an interface to ADC pins. 
This interface can be used to sample continuous voltages in the range of $0-3.3$V, and to convert them to discrete values. 
The ADC converts analog signals to 12 bit digital, covering a range from 0 to 4095. However, Micropython uses 16 bits to represent these values internally. 

The PWM is used to control the amount of time a signal is ON (duty rate) on an output defined GPIO. 
There are 12 independent PWM generators called slices (in the case of RP2040 there are only 8 PWM generators), each having two channels. 
The generators can be clocked from approximately 8Hz to 75MHz (8Hz to $62.5M$hz in the case of RP2040), at a machine frequency of 150MHz (125Mhz for RP2040). 
While the two channels of a slice will run at the same frequency, they can have a different duty rate.

The RP2 can communicate with peripherals using several standard interfaces such as: PWM, UART, SPI, I2C, I2S etc. 
In addition to these, the RP2 also allows us to define our own device interfaces by using PIO. 
The RP2350 has three PIO blocks (there are only two PIO blocks for RP2040), each of them contains four State Machines (SMs). 
Each SM can be used to manipulate the GPIO pins, and to transfer data. We can think of a SM as an independent processor able of running
up to a total of 32 instructions stored in a shared instruction memory.

Each SM has access to two FIFO registries (First In First Out) used to transmit and receive data, and each FIFO can be set to receive or transmit data independently. 
The SM has access to a shared bank of eight interrupt flags, which can be used for synchronization, to lock the SM, or for notifying the CPU. 
The first four interrupt flags are also exposed to the CPU and can trigger interrupts there. 
Also, each SM can control the GPIO pins by using four different sets of GPIO mappings: \verb|Input|, \verb|Output|, \verb|Set|, \verb|SideSet|. 
Each SM can have 0 to 32 \verb|Input| pins, 0 to 32 \verb|Output| pins, 0 to 5 \verb|Set| pins, and 0 to 5 \verb|SideSet| pins. 
The pins must be contiguous, but they can also overlap. 

In the \verb|Input| mapping we can define the input pin base, and all the other input pins must be contiguous counting up and modulo 32 wrapping.  
The \verb|Output| mapping works the same way, except that we also have to define how many pins we want to map as output. 
The SM can also control the direction of all pins, except of the input pins, and they can have overlapping mappings. 
The \verb|Set| and \verb|SideSet| mappings work similarly to the output mapping, with the exception that we can map a maximum of five GPIOs. 

A SM consists of a set of registers and a \verb|Control Logic|. 
The \verb|Program Counter| (PC) allows the SM to run independently from the others, and determines from where it should read the next instruction in the shared memory. 
The \verb|Control Logic| performs a basic increment function used for loops.
The \verb|Clock Divider| is used to run the SM at a frequency between 2 kHz and the 150MHz system clock (133MHz for the RP2040). 
The \verb|Clock Divider| can divide the system clock by 1 to 65536, with a fractional precision of 1 divided by 256. 
The \verb|In Shift| (ISR) and \verb|Out Shift| (OSR) registers are used for receiving (RX) and transmitting (TX) FIFO data, 
and the \verb|Scratch X| and \verb|Scratch Y| registers are used for temporary holding the data. Each register is 32 bits wide, and they can be 
configured to shift either left or right. 
The OSR can be used to shift data directly on the output pins. 
The ISR can be used to read data from input pins. This can be done 
by shifting the value from the pin into the ISR (we can also shift multiple pin values).  
The OSR and ISR are also used to push and pull data to the TX, and respectively RX FIFO.  
The shift registers can be also configured to automatically pull and push data once they are full. 
The FIFO is also aware of how much data has been pushed or pulled from it, and if the FIFO is full or empty is also detectable. 

The SM programs are written with a dedicated PIO assembler code consisting of 9 instructions: \verb|JMP|, \verb|WAIT|, \verb|IN|, \verb|OUT|, \verb|PUSH|, \verb|PULL|, \verb|MOV|, \verb|IRQ|, and \verb|SET|.
Also, Micropython has the included \verb|rp2| module acting as a wrapper for the PIO assembly language, such that the PIO assembly 
code is written as a function with the decorator \verb|@rp2.asm_pio()|.

Direct memory access (DMA) allows access to the memory independently of the CPU. 
The RP2350 microcontroller has 16 independent DMA (Direct Memory Access) channels and 4 separate IRQ (Interrupt Request) lines (the RP2040 only has 12 DMA channels and 2 IRQ lines).

The independent DMA channels can be used to execute a sequence of bus transfers in the following scenarios:

- Memory-to-peripheral: the DMA reads data from an array and writes to the peripheral's data FIFO.

- Peripheral-to-memory: the DMA reads data from the peripheral's data FIFO, and writes it to an array.

- Memory-to-memory: the DMA transfers data between two memory buffers.

With every clock cycle, the DMA controller can perform one read and one write, and the transfer size can be either 32, 16, or 8 bits (source and destination must have the same size).
Also, the DMA shares bandwidth evenly between the channels, executing a round-robin over all channels requesting data transfers.
Each channel has its own control and status registers, updating the process continuously.

\section{PIO-DMA wave function generator}

A wave function generator (WFG) is used to generate different types of signals (waveforms) for electronic circuits test purposes over a wide range of frequencies. 
The most common waveforms produced by the WFGs are: sine, square, ramp, and triangular. 

In order to build an RP2 controlled WFG we need to create a parallel port on the GPIO pins 0-7, 
and to repeatedly write the bits of the waveform signal to this port.  
In most cases a WFG with 8 bit resolution is enough. 
If necessary, one can easily modify this approach to achieve 10, or even 12 bits resolution, by simply changing a few lines of code. 

\begin{figure}[H]
\centering \includegraphics[width=3.5in]{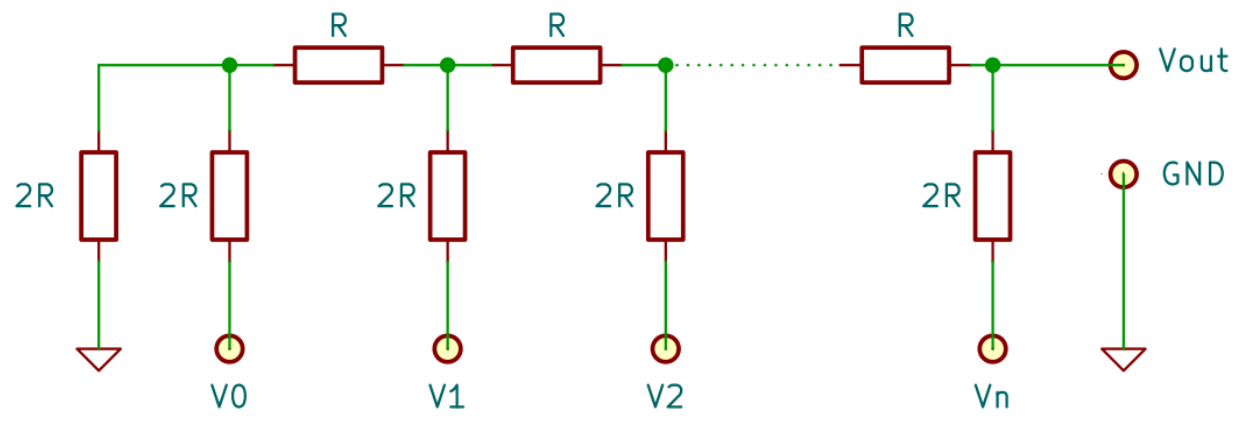}
\caption{R-2R ladder DAC.}
\end{figure}

Here we use $2^8=256$ integers to represent a period of a waveform. However, to actually generate the corresponding analog signal, we need to convert 
the parallel digital output into an analog signal using a DAC. Unfortunately the RP2 boards do not include a DAC, so we have to build one ourselves. 
Such a DAC can be built using the R-2R ladder network approach \cite{key-11}. As the name suggests, the R-2R uses just two resistive values, R and 2R. 
It can be easily shown that the equivalent resistance of the ladder is just R. 

To test this approach we first build an 8 bit R-2R ladder (Figure 1). The resistors used here are $\text{1K}\Omega$ and $\text{2K}\Omega$ respectively, with an error of $1\%$. 
The inputs to the circuit are the $V_0$ (GPIO0) to $V_7$ (GPIO7) potential values, which can be either HIGH (3.3V) or LOW (0V). 

Assuming that $m$ is an 8 bit integer, it can be written binary as:
\begin{equation}
m = \sum_{i=0}^7 b_i 2^i, \quad b_i \in \{0,1\}
\end{equation}
Thus, we can associate the bits $b_i$ with the $V_i$ voltages, such that for a digital input $m$, the output is:
\begin{equation}
V_{out} = V_{cc} \frac{m}{2^8} = 3.3 \frac{m}{256} V
\end{equation}

In most practical cases an 8 bit resolution is enough, if we take into account also the inherent signal smoothing due to the parasitic capacitance of the circuit. 
If necessary, one can easily modify this approach to achieve 10, or even 12 bits resolution, by simply changing a few lines of code. 
Also, in our case we would like to make two WFGs, which will require 16 output controlling GPIOs (Listing 1).

\begin{lstlisting}[language=Python, caption={Two channels 8 bit WFG.}]
import array, uctypes, rp2
from random import randint
from machine import Pin
from math import sin, exp, pi

def wfg_sig(sig):
    if sig == "sine":
        return bytes([int(128+127*sin(n*pi/128)) for n in range(256)])
    elif sig == "ramp":
        return bytes([n for n in range(256)])
    elif sig == "triangle":
        return bytes([abs(255-2*n) for n in range(256)])
    elif sig == "pulse":
        return bytes([int(255*exp(-(32*(n/256-0.5))**2)) for n in range(256)])
    elif sig == "square":
        return bytes([255*int(n/128) for n in range(256)])
    elif sig == "rand":
		return bytes([randint(0,255) for n in range(256)])
    
array0 = array.array('B',[0]*512)
addr0 = uctypes.addressof(array0)
aligned0 = (addr0 + 255) & ~255
offset0 = aligned0 - addr0
wave0 = memoryview(array0)[offset0 : offset0 + 256]

dma0 = rp2.DMA()
ctrl0 = dma0.pack_ctrl(size=0,
                       inc_read=True,
                       inc_write=False,
                       treq_sel=0,
                       ring_size=8,
                       ring_sel=False)

@rp2.asm_pio(out_init=(rp2.PIO.OUT_LOW,)*8,
             out_shiftdir=rp2.PIO.SHIFT_LEFT,
             autopull=True,
             pull_thresh=8)
def pio0():
    out(pins, 8)
    
sm0 = rp2.StateMachine(0, pio0, freq=256*1000, out_base=Pin(0))
def config_wave0(sig="none", freq=1000):
    sm0.active(0)
    dma0.active(0)
    if sig != "none":
        sm0.init(pio0, freq=256*freq, out_base=Pin(0))        
        wave0[:] = wfg_sig(sig)
        dma0.config(read=aligned0,
                    write=sm0,
                    count=4_294_967_295,
                    ctrl=ctrl0,
                    trigger=True)    
        dma0.active(1)
        sm0.active(1)

array1 = array.array('B',[0]*512)
addr1 = uctypes.addressof(array1)
aligned1 = (addr1 + 255) & ~255
offset1 = aligned1 - addr1
wave1 = memoryview(array1)[offset1 : offset1 + 256]

dma1 = rp2.DMA()
ctrl1 = dma1.pack_ctrl(size=0,
                       inc_read=True,
                       inc_write=False,
                       treq_sel=8,
                       ring_size=8,
                       ring_sel=False)

@rp2.asm_pio(out_init=(rp2.PIO.OUT_LOW,)*8,
             out_shiftdir=rp2.PIO.SHIFT_LEFT,
             autopull=True,
             pull_thresh=8)
def pio1():
    out(pins, 8)
    
sm1 = rp2.StateMachine(4, pio1, freq=256*1000, out_base=Pin(8))

def config_wave1(sig="none", freq=1000):
    sm1.active(0)
    dma1.active(0)
    if sig != "none":
        sm1.init(pio1, freq=256*freq, out_base=Pin(8))        
        wave1[:] = wfg_sig(sig)
        dma1.config(read=aligned1,
                    write=sm1,
                    count=4_294_967_295,
                    ctrl=ctrl1,
                    trigger=True)
        dma1.active(1)
        sm1.active(1)
\end{lstlisting}

We start with a function that returns the byte arrays for the waveforms we would like to generate (sine, ramp, triangular, pulse, square and rand). 
The \verb|pulse| is defined as a narrow Gaussian centered on the 256 bytes array. The \verb|rand| option uses the \verb|randint()| function to  
generate random bytes, and in this case it does not need to be seeded. 
The rest are well known functions. 

During the signal generation process, the array must be stored as a memory-aligned byte array, since the standard \verb|bytearray()| constructor does not guarantee contiguous data storage.
Therefore we allocate a double sized array and slice it using an offset to wrap an aligned memory region. Here, \verb|aligned0| is the 
address of the aligned data stored in \verb|array0|, and \verb|wave0| is the memory view of this data array. 

Next step is to define the DMA channel that will push the data to the PIO state machine, which will output the byte values in parallel to an 8 bit R-2R DAC. 
The \verb|dma0 = rp2.DMA()| claims and initializes an unused DMA channel automatically. Then, the \verb|dma0.pack_ctrl(*)| method is used to pack 
the control register configuration fields into a single integer \verb|ctrl0|. Here, the function arguments have the following meaning:

\begin{itemize}
\item \verb|size=0|: sets the transfer size to 1 byte. 

\item \verb|inc_read=True|: increments the read pointer address after every single transfer (iterates through the array storing the waveform bytes). 

\item \verb|inc_write=False|: means that DMA writes continuously at the same address pointer. 

\item \verb|treq_sel=0|: the data request (DREQ) used to pace the transfer based on the PIO peripheral readiness. 

\item \verb|ring_size=8|: address wrapping sizing exponent (in our case the address wraps at a $2^8=256$ bytes boundary of \verb|data0|).

\item \verb|ring_sel=False|: dictates which pointer wraps (False applies \verb|ring_size| to the read address; True applies it to the write address).
\end{itemize}

This step is followed by the configuration of the PIO state machine which auto pulls the data, and its role 
is to simply output the byte values (starting from GPIO 0 up to GPIO 7) of the wave function requested by the \verb|config_wave0| function. 
The \verb|dma0.config(*)| function is used to apply the configuration to the DMA channel registers with the following parameters: 

\begin{itemize}

\item \verb|ctrl=ctrl0|: this is the previously defined configuration integer.

\item \verb|count=4_294_967_295|: means the channel is configured for 4,294,967,295 transfers (maximum value of an unsigned 32-bit integer).

\item \verb|trigger=True|: starts the DMA transfer immediately.

\item \verb|write=sm|: means that the DMA channel will write at the defined state machine (sm). 

\item \verb|read=aligned0|: is the starting address of the aligned data buffer of the waveform (\verb|data0| in our case). 
\end{itemize}

\begin{figure}[H]
\centering \includegraphics[width=6.4in]{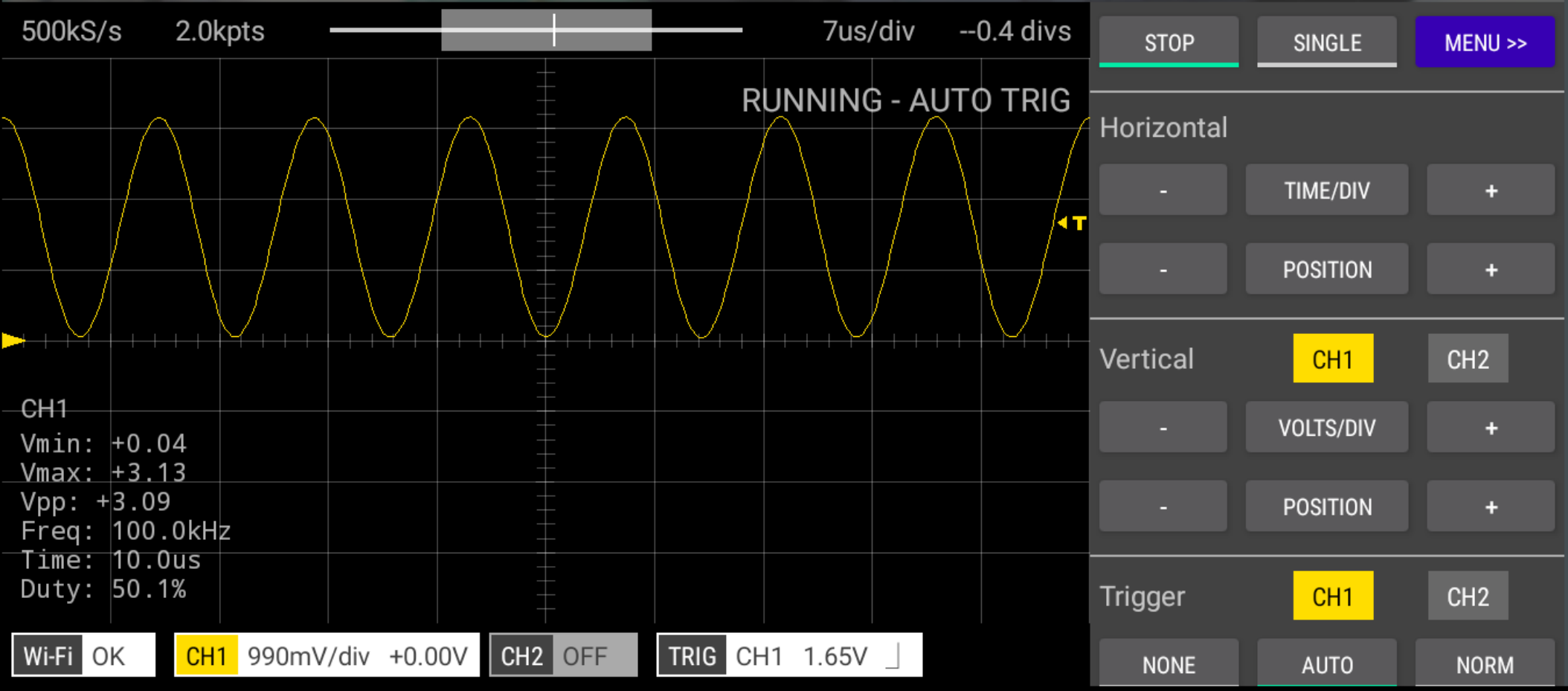}
\centering \includegraphics[width=6.4in]{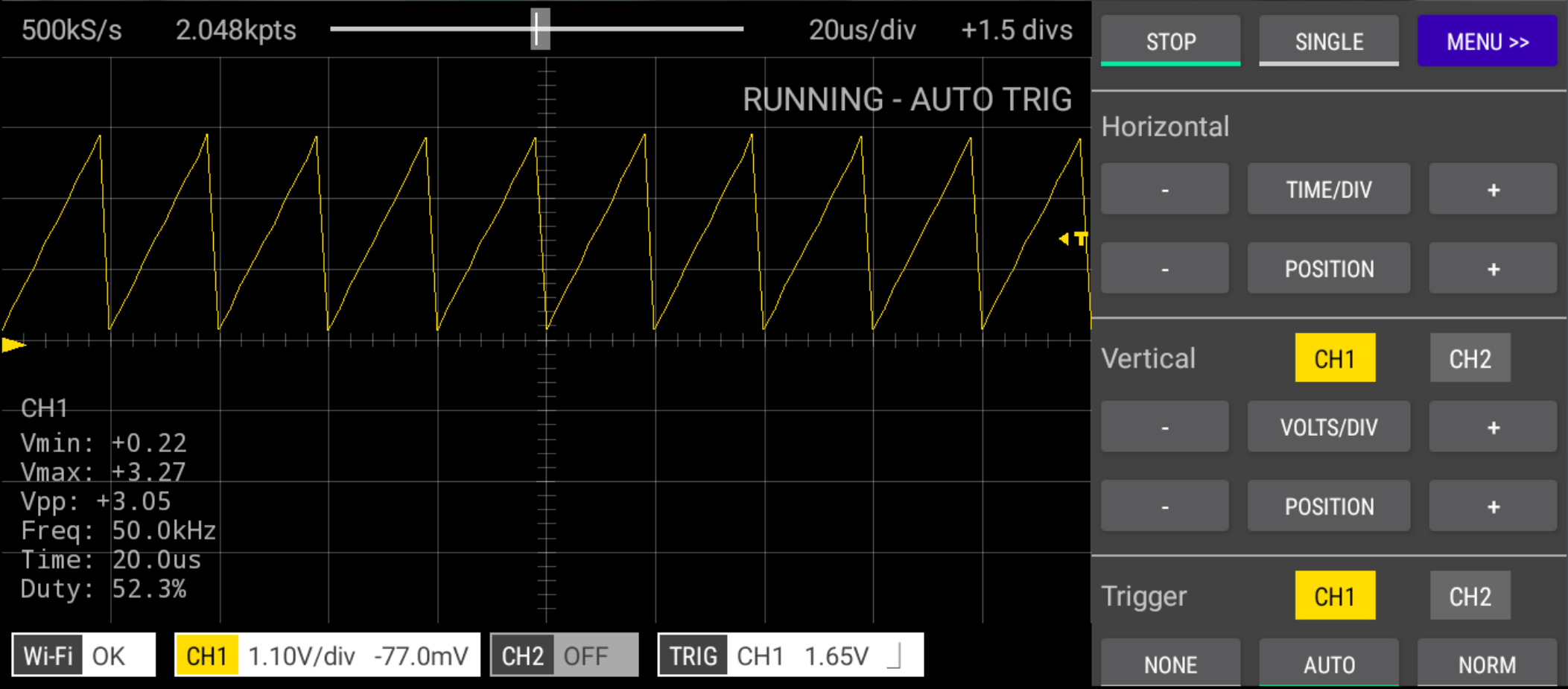}
\caption{Testing the 8 bit WFG: 100 KHz sine signal, and a 50 KHz ramp signal.}
\end{figure}

To define a second WFG channel, we just duplicate the code and replace the 0 suffix with 1 on all the variables and function names. 
We save the code as \verb|wfg.py| on the RP2. 

In order to test the WFG code, we can run the simple test in Listing 2, and use an oscilloscope to visualize the results. 
Of course, this test would require to actually implement the R-2R DAC also, which for testing purposes it can be done on a breadboard.  
An example is shown in Figure 2, where we generate a sine wavefunction at 100KHz on channel 0, and a 50KHz ramp signal on channel 1. 
The results look quite acceptable considering that for these testing purposes the R-2R DAC circuit was simply built on a breadboard.

\begin{lstlisting}[language=Python, caption={Testing the 8 bit WFG.}]
from wfg import *

config_wave0("sine", 100_000)
config_wave1("ramp", 50_000)
\end{lstlisting}

We should note that \verb|count=4_294_967_295| in \verb|dma.config| behaves differently on RP2305 and RP2040. 
On RP2350 this maximum value sets a continuous transfer mode, by safely configuring the hardware to wrap around and instantly reload.
On RP2040 the transfer effectively ends when \verb|count=4_294_967_295|, and the process needs a restart.

\section{DMA ADC oscilloscope}

RP2 has an internal ADC allowing for 500KS/s (kilo samples per second) acquisition using a 48MHz clock. 
The digital controller managing ADC provides the following programmable functionality:
\begin{itemize}
\item one-shot or free-running capture mode.

\item DMA interface for DMA FIFO sampling.

\item pacing timer for setting the sample rate. 

\item round-robin sampling of multiple channels.

\item optional right-shift to 8 bits in free-running capture mode.

\end{itemize}

RP2 has four ADC channels (GPIO26, GPIO27, GPIO28, GPIO29), however only the first three are accessible to the user, the forth one is linked to the internal temperature sensor. 

According to the official documentation of the RP2 boards, 96 cycles are required for an ADC conversion (both RP23050 and RP2040) \cite{key-7}, \cite{key-8}. 
Since the standard clock frequency is set at 48MHz, a conversion takes $2\mu s$ per sample, and therefore we can have a maximum sample rate of 500KS/s. 
The sample rate can be reduced by increasing the clock frequency divider to a positive value (Listing 3), which initially is set to \verb|div_val=0| for for the maximum value of 500KS/s. 
This gives us the possibility of controlling the sample rate over a large range of integer values: $1875,3750,7500,15000,30000,60000,120000,240000,480000,500000$. 

\begin{lstlisting}[language=Python, caption={ADC clock frequency divider.}]
if rate == 500_000:
    div_val = 0
else:
    div_val = int((48_000_000 // rate) - 1) << 8
\end{lstlisting}

The ADC controller can be used to sample multiple inputs in an interleaved fashion. However, all the ADC channels share the same bandwidth, which divides the maximum 500KS/s sample rate 
between them. 
In order to be able to maintain a high sampling rate, here we prefer to use a different approach, by quickly alternating the sampling between two channels (GPIO26, GPIO27), such that our oscilloscope 
will operate with two input channels at maximum sampling rate. This scheme can be easily modified to include a third channel if necessary, 
however an additional channel will also require an additional OP-AMP, which will complicate the frontend electronics implementation. 

The ADC samples can be taken from the RESULT register or stored into the local FIFO. The FCS register is used to control the FIFO operation.
If FCS.EN is set to 1, then ADC writes the result to the FIFO. The DMA can read this sample from the FIFO when notified by the DREQ (data request) signal. 
When the FIFO is full, the sticky error flag FCS.OVER is set, and the current FIFO contents do not change (any additional conversions will be lost). 
DMA can be used to fetch the samples by performing a memory mapped read on the FIFO registry, and this process is paced by the \verb|ADC_DREQ| signal. 

More details about the optimal DMA configuration can be found in the official documentation, here we prefer to discuss directly on our implementation code. 
Since the official documentation is quite useful for understanding the process, we strongly recommend to consult it before jumping into the coding section \cite{key-7}, \cite{key-8}. 
Our DMA ADC implementation is given in Listing 4.

\begin{lstlisting}[language=Python, caption={DMA ADC data acquisition.}]
import sys, uctypes, array, rp2
from machine import ADC, Pin, mem32, freq
from time import sleep

freq(240_000_000); sleep(0.1)

if "RP2350" in sys.implementation._machine.upper():
    ADC_BASE = 0x400a0000
    CLK_ADC_CTRL = 0x4001006c
    TREQ_SEL = 48
else:
    ADC_BASE = 0x4004c000
    CLK_ADC_CTRL = 0x40008060
    TREQ_SEL = 36        
    
mem32[CLK_ADC_CTRL] = 0x820; sleep(0.1)

Pin(23, Pin.OUT).value(1)

ADC(Pin(26, Pin.IN, pull=None))
ADC(Pin(27, Pin.IN, pull=None))

dma2 = rp2.DMA()

def DMAadc(adc_chan=0, samples=2048, rate=480_000):
    samples = samples + 256
    ADC_CS_ADDR     = ADC_BASE + 0x00
    ADC_RESULT_ADDR = ADC_BASE + 0x04
    ADC_FCS_ADDR    = ADC_BASE + 0x08
    ADC_FIFO_ADDR   = ADC_BASE + 0x0c
    ADC_DIV_ADDR    = ADC_BASE + 0x10       
    capture_buffer = array.array('H', [0] * samples)
    buffer_memory_address = uctypes.addressof(capture_buffer)       
    try:                
        mem32[ADC_FCS_ADDR] = 0x0
        while not (mem32[ADC_FCS_ADDR] & (1 << 8)):
            g = mem32[ADC_FIFO_ADDR]
        mem32[ADC_FCS_ADDR] |= (1 << 11) | (1 << 10)
        mem32[ADC_FCS_ADDR] = (1 << 3) | (1 << 0) | (1 << 24)
        mem32[ADC_CS_ADDR] = 0x0        
        mem32[ADC_CS_ADDR] = (adc_chan << 12) | (1 << 0)        
        if rate == 500_000:
            div_val = 0
        else:
            div_val = int((48_000_000 // rate) - 1) << 8        
        mem32[ADC_DIV_ADDR] = div_val
        ctrl = dma2.pack_ctrl(size=1,
                              inc_read=False,
                              inc_write=True,
                              treq_sel=TREQ_SEL)
        dma2.config(read=ADC_FIFO_ADDR,
                    write=buffer_memory_address,
                    count=samples,ctrl=ctrl)        
        dma2.active(1)
        mem32[ADC_CS_ADDR] |= (1 << 3)
        while dma2.active():
            pass
        mem32[ADC_CS_ADDR] &= ~(1 << 3)
        dma2.active(0)        
        return memoryview(capture_buffer)
    except Exception as e:
        print("Error:", e)
\end{lstlisting}
 
We should point out that there are some differences between the two boards RP20350 and RP2040 \cite{key-7}, \cite{key-8}. 
First, the ADC registers addresses are different, with the newer RP2350 as \verb|ADC_BASE = 0x400a0000|, 
while the former RP2040 version as \verb|0x4004c000|. The second difference is in the address of the clock register, 
which is \verb|CLK_ADC_CTRL = 0x4001006c| for RP2350, and \verb|0x40008060| for RP2040. 
The third difference is the required value \verb|TREQ_SEL| value for peripheral data request (DREQ) signal, which is used to pace the DMA transfers. 
This value is 48 for the RP2350, and 36 for RP2040. Other than these differences, the code is the same for both RP2 boards. 

We should emphasize that the code is overclocking the boards to 240MHz (from 150MHz for RP2350, and respectively 133MHz for RP2040). 
We chose the 240MHz value because it provides the same integer frequency conversion constant $c=240/48=5$ for both boards. 
This choice is a compromise, such that the software has the same performance on both boards, and one can "plug and play" any of the boards. 
The maximum safe overclocking frequency is 300MHz for RP2350 ($c=6.25$), and 266MHz for RP2040 ($c=5.541666667$). 

The code sets the \verb|ADC_BASE|, \verb|CLK_ADC_CTRL| and \verb|TREQ_SEL| corresponding values for the board, 
and selects the system clock as the clock source \verb|mem32[CLK_ADC_CTRL] = 0x820| (changing the ADC clock source is not necessary if the system is not overclocked). 
The \verb|dma2 = rp2.DMA()| selects the next available DMA channel. We denoted this as \verb|dma2| since the \verb|dma0| and \verb|dma1| channels are used by the previously described WFG implementation. 
The \verb|DMAadc(adc_chan=0, samples=2048, rate=480_000)| function performs the data acquisition with the following parameters:

\begin{itemize}

\item  \verb|adc_chan|: the ADC channel used, 0 for ADC0 (GPIO26), 1 for ADC1 (GPIO27).

\item \verb|samples|: the number of samples requested. 

\item \verb|rate|: the ADC sampling rate. 

\end{itemize}

We are also increasing the requested number of samples with 256 extra samples, 
because later we will use these extra samples to synchronize the channels for triggering and visualization purposes in the oscilloscope user interface.
Next we set the address of the registries used, and we initialize the buffer for data acquisition. 
The code continues by clearing any residual values in FIFO (from a previous run) and resets control settings. 
Then, using \verb|mem32[ADC_CS_ADDR]| we wake the ADC hardware block (by setting bit 0 to 1), and we select the ADC channel (by setting bit 12 with the \verb|adc_chan| value).
As discussed before, the next step is to set the desired sample rate: \verb|mem32[ADC_DIV_ADDR] = div_val|.

The \verb|dma2.config(*)| is used to configure the DMA channel with the structured \verb|ctrl| integer constant. Here \verb|size=1| means 16 bit values, 
\verb|inc_read=False| tells DMA to read from the same \verb|read=ADC_FIFO_ADDR| address, and \verb|inc_write=True| to write the result at buffer address \verb|write=buffer_memory_address|, 
while \verb|treq_sel=TREQ_SEL| performs the peripheral data request. The process is then configured to the continuous acquisition of \verb|count=samples| number of samples. 
This step is followed by setting the DMA active \verb|dma2.active(1)|, and by starting the acquisition process with setting the bit 3 of the \verb|ADC_CS_ADDR| register to 1. 
The acquisition process is then executed and finally stopped with \verb|dma2.active(0)| and setting the bit 3 of the \verb|ADC_CS_ADDR| register to 0, once it has captured the number of requested samples. 
Finally, the captured data buffer is returned, and the values are printed for a quick visual inspection. 

We save this code as \verb|adc.py|, and for testing purposes we can simply run the script from Listing 5, which takes 1024 samples at an acquisition rate of 48MS/s, and prints the results 
on the console. The test returns an array of 1024 values which are then printed on the console. These are rather small and noisy values since we didn't had any input signal, 
and we basically just measured the ADC noise, because the ADC input pins are configured without a pulling HIGH or LOW a resistor.

\begin{lstlisting}[language=Python, caption={Testing the DMA ADC data acquisition.}]
from adc import *

data = DMAadc(adc_chan=0,samples=1024,rate=48_000_000)
for i in range(len(data)):
    print(data[i],end=",")
    
6,2,4,5,5,3,3,5,5,3,3,6,5,3,3,4,7,4,3,5,6,3,3,4,5,...
\end{lstlisting}

\section{RP2 software package}

We can now proceed with packaging the software that will run on the RP2 board. 
This on-board software will communicate with the controlling software running on the browser of a computer. 
The browser software will be described in a later section, here we just focus on the software that will run on the RP2 board. 

RP2 boards automatically run the \verb|main.py| program when they are powered through the micro-USB port. The \verb|main.py| code is shown in Listing 6, 
and it can run two WFG channels and two ADC channels. Once \verb|main.py| runs on RP2, it will 
wait for a connection request from the Javascript software that runs on the browser.

\begin{lstlisting}[language=Python, caption={The oscilloscope "main.py" program.}]
import json
from gc import collect
from machine import PWM
from adc import *
from wfg import *

pwm = [PWM(Pin(16)), PWM(Pin(18))]

dN = 2048
@micropython.viper
def stream_data(data):
    N = len(data) 
    for i in range(0, N, dN):
        sys.stdout.write(",".join(map(str, data[i:i + dN])))
        if i + dN < N:
            sys.stdout.write(",")
    sys.stdout.write("\n")    

while True:
    try:        
        ctrl = json.loads(sys.stdin.readline().strip())
        
        if ctrl["wfg"][0]["as"] == 1:
            config_wave0(sig=ctrl["wfg"][0]["sig"], 
                         freq=ctrl["wfg"][0]["freq"])
                         
        if ctrl["wfg"][1]["as"] == 1:
            config_wave1(sig=ctrl["wfg"][1]["sig"], 
                         freq=ctrl["wfg"][1]["freq"])
        if ctrl["pwm"][0]["cs"] == 1:
            pwm[0].freq(ctrl["pwm"][0]["freq"])
            pwm[0].duty_u16(int(655.35*ctrl["pwm"][0]["duty"]))
            
        if ctrl["pwm"][1]["cs"] == 1:
            pwm[1].freq(ctrl["pwm"][1]["freq"])
            pwm[1].duty_u16(int(655.35*ctrl["pwm"][1]["duty"]))
            
        collect()            
        
        stream_data(DMAadc(adc_chan=ctrl["adc"][ctrl["idx"]]["chan"],
                           samples=ctrl["adc"][ctrl["idx"]]["samples"],
                           rate=ctrl["adc"][ctrl["idx"]]["rate"]))                
    except Exception as e:
        print(f"Error:{e}")
\end{lstlisting}

Here we have also defined two PWM channels on GPIO16 and GPIO18, for providing the PWM functionality, and they will be controlled directly 
from the software interface running on the browser. 

The \verb|main.py| program consists of an infinite \verb|while| loop where the interaction 
between the RP2 board and the host computer browser takes place. 
Here, the \verb|ctrl| json dictionary is received from the Javascript software running on the host computer browser. 
The structure of the json is shown in Listing 7, and its content provides the control commands for the RP2 board, with 
the user requested settings for ADC, PWM and WFG.

\begin{lstlisting}[language=Python, caption={The ctrl json used for communication.}]
ctrl = {"idx":0,
        "adc": [{"samples":0,"chan":0,"rate":0},
                {"samples":0,"chan":1,"rate":0],
        "pwm": [{"freq":0,"duty":0,"cs":0},
                {"freq":0,"duty":0,"cs":0}],                   
        "wfg": [{"freq":0,"sig":"none","s":0},
                {"freq":0,"sig":"none","s":0}]}
\end{lstlisting}

The \verb|"idx"| field can be 0 or 1, and is only used on the browser side to identify the oscilloscope channel. 
The \verb|"adc"| field is used to communicate the ADC parameters: number of samples, channel (0,1) and the sampling rate. 
The \verb|"pwm"| field provides the parameters (frequency and duty) for the two PWM channels. Here, \verb|"cs"| is an ON/OFF switch variable, 
if \verb|cs=1| then RP2 must set the PWM channel ON with the frequency and duty provided, otherwise the PWM channel must be set OFF.
The \verb|wfg| field provides the frequency and type signal parameters for the two WFG channels. Again, \verb|"s"| is an ON/OFF switch variable, if \verb|s=1| then RP2 must set ON the WFG channel 
with the parameters provided, otherwise the WFG channel must be set to OFF.

In the next step we call the \verb|collect()| function to free memory for the data acquisition process. 
This step will clean up and create empty contiguous areas of memory that are required by the contiguous buffers used in the ADC process. 

Finally, the \verb|DMAadc(*)| function is called with the provided parameters, and after the data is collected it is sent to the web browser through the micro-USB interface, using the 
\verb|stream_data(data)| function. This function implements a streaming process where maximum 2048 data values are sent in one step of the iteration process, until all data is sent to the web browser. 
This way the process is controlled by the software running on the web browser, and runs continuously until the RP2 is disconnected from the micro-USB port. 
We also notice that the \verb|@micropython.viper| decorator used here compiles the function directly into native machine code (ARM Thumb assembly), 
allowing the loops to run up to 10 times faster than the standard interpreted Micropython code.

The code can be also compiled using the official Micropython cross-compiler, \verb|mpy-cross|, which converts the python source files into binary bytecode files. 
This saves RAM and storage space, such that the code loads faster and with less memory overhead. 

\section{Oscilloscope frontend}

The role of the oscilloscope frontend is to provide protection for the input ADC pins, which can only accept a voltage in the range of $0-3.3$V, and also to safely deliver  
the output signals generated by the WFG and PWM. 
Therefore, the frontend discussed here is minimal, and its purpose is mostly to protect the RP2 board from 
harmful voltages and currents. Also, the frontend should provide the required basic functionality to the whole system, such as switching between AC and DC input signals, 
or between the WFG and PWM output signals. 
The schematics of the frontend is shown in Figure 3.

\begin{figure}[H]
\centering \includegraphics[width=6.5in]{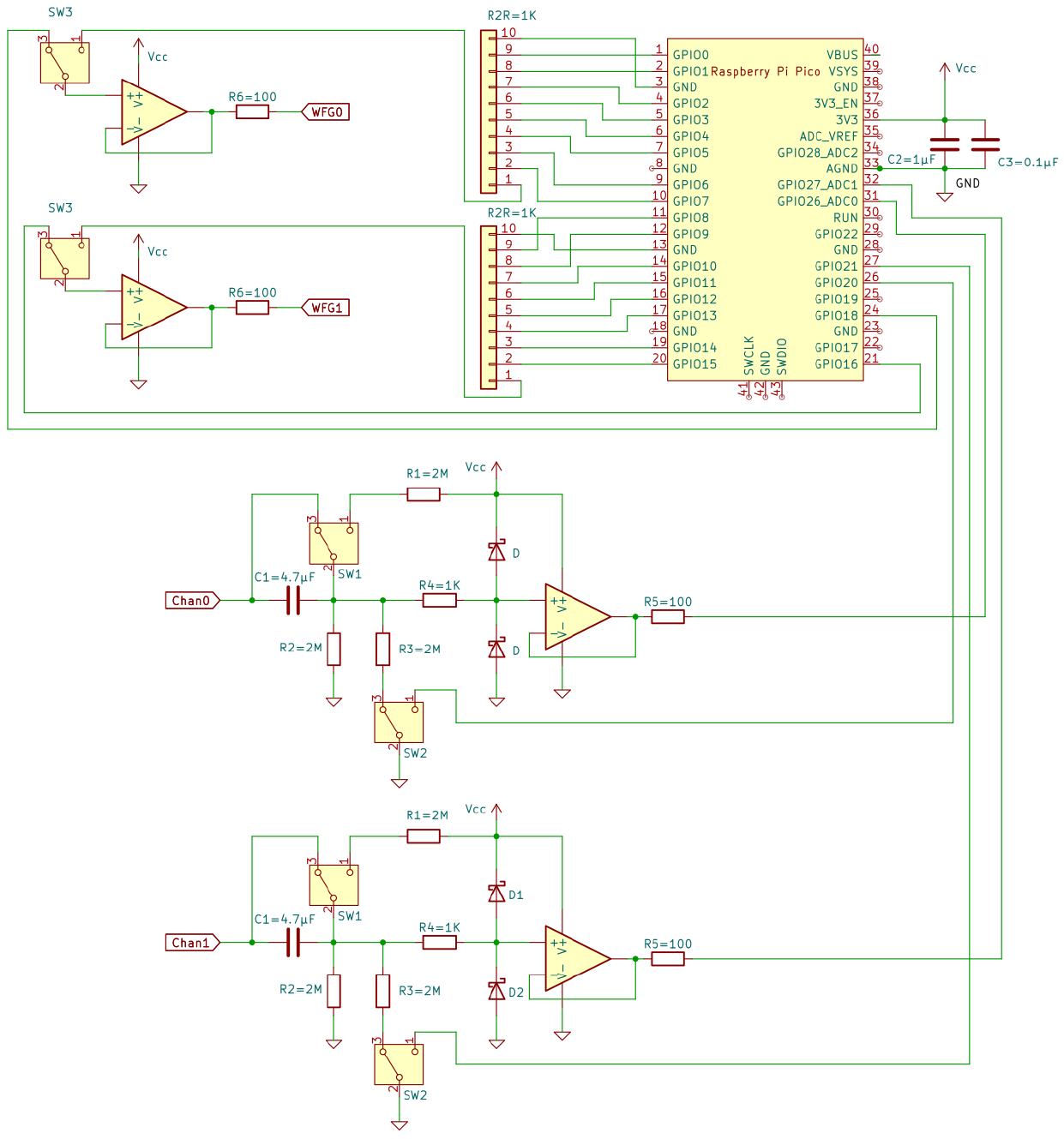}
\caption{Oscilloscope electronics frontend.}
\end{figure}

Here we use MCP6022 from Microchip Technology Inc. \cite{key-12}, which are cost effective dual operational amplifiers, offering rail-to-rail input and output, and 
high performance capabilities: wide bandwidth (10 MHz), low noise ($8.7$nV/$\sqrt{Hz}$), slew rate ($7.0$V/$\mu s$), low input offset voltage and low distortion ($0.00053\%$). 

The OP-AMPS are powered from the $3.3$V source on the RP2 board (pin 36), and are configured as non-inverting buffers, 
since their role is mostly to provide a high input impedance and a low output impedance. The capacitors C2=$1\mu$F and C3=$0.1\mu$F are used to filter low/high frequency 
ripples in the power supply (just in case they are present). 

A small resistor $\text{R5=R6=100} \Omega$ is placed at the output of the OP-AMPS in order to avoid the capacitive loads problem and to improve the stability, 
since the ADC pins have a small input sampling capacitance of approximately 1 pF. This resistor will decrease a bit the OP-AMP bandwidth from the initial 10 MHz, 
however it will not have a significant effect in our case. 

The two input switches, SW1 and SW2, are practically a single, double pole double throw (DPDT), switch with six terminals that controls the two separate circuits at the same time. 
If the poles are connected to position 3 then the resistor $\text{R1=2M}\Omega$ is not connected to the input, 
and the input coupling capacitor $\text{C1=4.7}\mu$F is shorted, such that the input resistor $\text{R2=2M}\Omega$ is in parallel with $\text{R3=2M}\Omega$, 
resulting into an equivalent input impedance of $\text{R}=\text{1M}\Omega$. This is the switch position required for DC input signals. 
On the contrary, for the input AC signals, the poles should be connected to position 1 such that the input impedance, which in this case is still $\text{R}=\text{1M}\Omega$, 
is dictated only by $\text{R1=2M}\Omega$ and $\text{R2=2M}\Omega$. 
The resistor $\text{R4=1K}\Omega$ is added to protect the OP-AMP against larger input currents. 
Similarly, the two Schottky diodes (BAT85), D1 and D2, are added to clamp the input signal voltages to $0-3.3$V. These are very fast switching diodes that do not distort the input signal, 
but constrain the input signal to a safe range. The output of these two OP-AMPS controlling the oscilloscope input are connected to the input ADC pins 31 and 32 of the RP2 board. 
Also, the pin 1 of the SW2 switches is connected to GPIO 26 and respectively GPIO 27, such that we can easily detect the GND state of the switch directly from software. 

\begin{figure}[H]
\centering \includegraphics[width=4in]{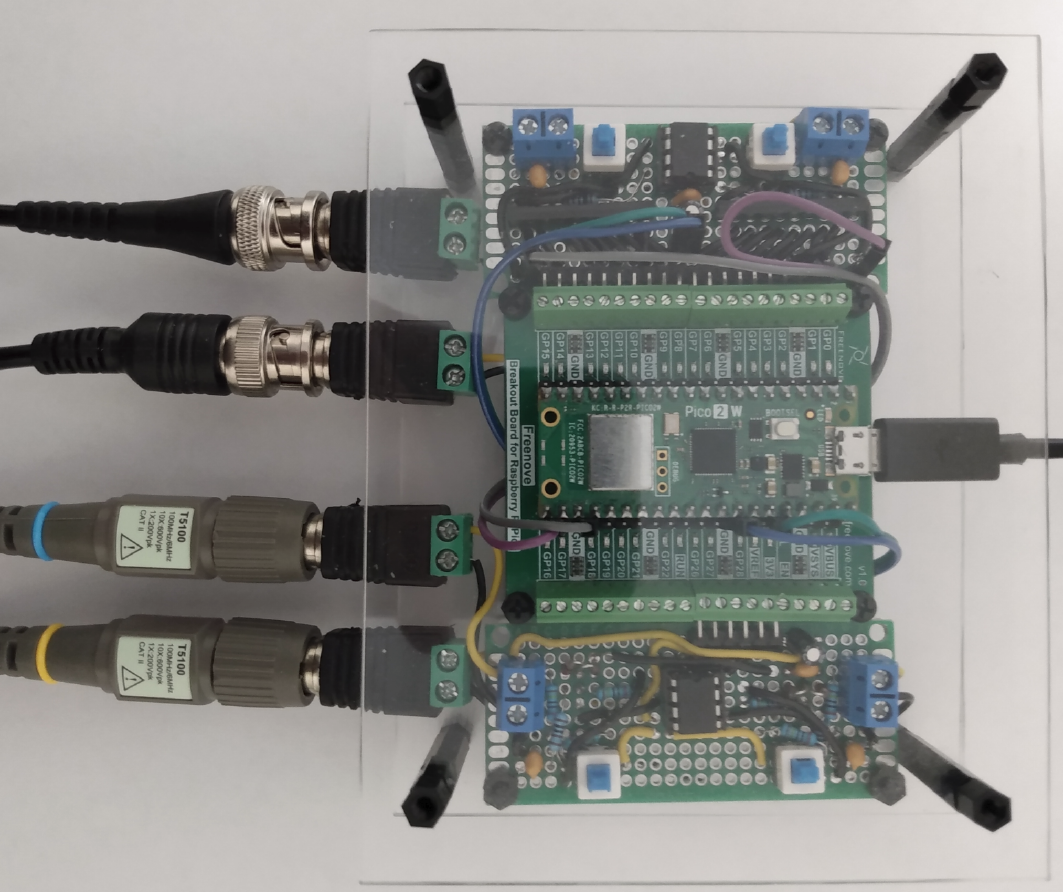}
\caption{The O-SCOPE.}
\end{figure}

On the other side, we have two OP-AMP buffers configured for the WFG and PWM output. First the WFG digital signal is processed by the R-2R networks, which are compactly represented here 
since we use an integrated R-2R circuit ladder (Bourns 4310R-101-102) where pin 10 is the GND pin, pins 9(LSB)-2(MSB) are the input pins, and pin 1 is the output. 
The output of the R-2R ladder is then connected to the pin 1 of a single pole double throw (SPDT) switch, which is connected to the input of the OP-AMP buffer. 
The pin 3 of the SPDT switch is also connected to the GPIO 16 (GPIO 17) which provides the PWM output signal. Thus, the SPDT switch can be used to select between the WFG and PWM output signals. 

The oscilloscope is enclosed into a small acrylic box ($12\text{cm} \times 5\text{cm} \times 2.5\text{cm}$), where we also attached the necessary BNC input/output connectors. 
Since the input impedance of the channels is $\text{R}=1\text{M}\Omega$ we can actually use true oscilloscope input probes as shown in Figure 4.

\section{The web browser user interface}

In the previous section we have shown how to build the distribution software running on the RP2 board, however to be able to make use of it we need to 
build a user interface running on a computer. Here we use the Web Serial API to establish the communication between the RP2 board and the browser running on the host computer \cite{key-6}. 
Also, we build a simple user interface for controlling the oscilloscope and to visualize the measurements. 
Currently Chrome is one of the Web Serial capable browsers, and this is what we use in our development. 
As mentioned before, the development uses only vanilla Javascript and HTML, and it does not have any dependencies. This allows us to run the web user interface by simply 
installing it on a free GitHub Pages account, such that it is accessible everywhere \cite{key-13}. 

The Javascript code is quite complex and long. Therefore, here we will just present the process of establishing the connection between the browser and the RP2 board (Listing 8), 
and some of the computation processes, such as the Fast Fourier Transform (FFT) of the signal (Listing 9). 

\begin{lstlisting}[caption={Javascript code used for communication.}]
const nsamples = 8192; 
const offset = 256; 
let port, reader, writer; 

const connectBtn = document.getElementById('connectBtn');
connectBtn.addEventListener('click', async () => {
    try {
        port = await navigator.serial.requestPort();   
        await port.open({baudRate: 921600, bufferSize: 131072});
        connectBtn.disabled = true;
        await sendCommand(JSON.stringify(ctrl));   
        await listenToPico();
    } catch (err) {
        console.log(err);
    }
});

async function sendCommand(ctrl) {
    if (!port.writable) {
        return;
    }
    const textEncoder = new TextEncoder();
    writer = port.writable.getWriter();
    await writer.write(textEncoder.encode(ctrl + "\n"));
    writer.releaseLock();      
}

async function listenToPico() {
    if (!port.readable) {
        return;
    }
    const textDecoder = new TextDecoderStream();
    const readableStreamClosed = port.readable.pipeTo(textDecoder.writable);
    reader = textDecoder.readable.getReader();  
          
    let buffer = "";    
    try {
        while (true) {       
            const { value, done } = await reader.read();
            if (done) {
                reader.releaseLock();
                break;
            }
            buffer += value;           
            const lines = buffer.split("\n");            
            for (let i = 0; i < lines.length - 1; i++) {
                const jsonString = lines[i].trim();
                if (jsonString) {
                    try {
                        const parsedArray = jsonString.split(",").map(Number);
                        if (nsamples == parsedArray.length - offset) {
                            processData(parsedArray);
                        }
                    } catch (e) {
                        console.error("JSON parsing error", e);
                    }
                }
            }            
            buffer = lines[lines.length - 1];
        }
    } catch (error) {
        console.error("Read error: ", error);
    } 
    finally {
        reader.releaseLock();
    }
}

\end{lstlisting}

In the user interface the connection is established using a HTML button element (\verb|'connectBtn'| ) associated with an event listener as shown in Listing 8.
Once the connection is established the browser sends the \verb|ctrl| json as a string using the \verb|sendCommand(ctrl)| function 
to the RP2, which then responds by sending back the requested data (\verb|nsampes=8192| plus the \verb|offset=256| for synchronization visualization purposes). 
The data sent by the RP2 is received and parsed by the \verb|listenToPico()| function, which calls the 
\verb|processData()| function to process the received data. 

Once the data is received and parsed, the signal is further processed to extract some relevant statistical values (maximum, minimum, average,...), and to calculate the FFT, 
which converts a time-domain signal into its frequency-domain components. The FFT is based on an efficient recursive implementation of the Cooley-Tukey algorithm \cite{key-14}, 
and requires an input array length that is a power of 2 (Listing 9). 

\begin{lstlisting}[caption={FFT implementation.}]
function FFT(signal) {
    const N = signal.length;
    if (N < 2) return signal.map(val => [val, 0]);
    const N2 = Math.trunc(N/2);
    const evens = FFT(signal.filter((_, i) => i % 2 === 0));
    const odds  = FFT(signal.filter((_, i) => i % 2 !== 0));    
    const result = new Array(N);
    for (let k = 0; k < N2; k++) {    
        const theta = -2 * Math.PI * k / N;
        const wr = Math.cos(theta); 
        const wi = Math.sin(theta); 
        const tr = wr * odds[k][0] - wi * odds[k][1];
        const ti = wr * odds[k][1] + wi * odds[k][0];
        result[k] = [evens[k][0] + tr, evens[k][1] + ti];
        result[k + N2] = [evens[k][0] - tr, evens[k][1] - ti];        
    }
    return result;
}
\end{lstlisting}

The user interface also provides the following features:

\begin{itemize}

\item USB connection button.

\item Channel type selection, DC or AC, and WFG or PWM output selection.

\item Signal zoom selection in a range from 1 to 6.

\item Sampling rate selection in the range: 1875, 3750, 7500, 15000, 30000, 60000, 120000, 240000, 480000, 500000 multiplied by the system clock constant $c=5$ 
(the maximum sampling rate is actually $2.5$MS/s).  

\item Trigger selection: channel, rising or falling signal, auto or single. 

\item Vertical cursor setting for the trigger threshold value. 
      Two horizontal cursors for measuring both time and voltage signal values, and the difference between these values at different cursor positions.

\item Settings for the two WFG channels: signal type (sine, ramp, triangular, pulse, square, random), and frequency (from 18 Hz to 250KHz). 

\item Settings for the two PWM channels: frequency (from 18 Hz to 250KHz), and duty (from $0\%$ to $100\%$). 

\item Automatic maximum peak detection in the FFT windows.

\item Automatic scaling of the measured signals in the visualization windows. 

\item Real time interactive visualization of the signals and their FFTs.

\end{itemize}

In addition, the user interface provides two buttons STOP/RUN and SAVE DATA. The STOP/RUN button can be used to freeze the 
acquisition and visualization process. This is useful for visually inspecting fast changing signals. Also, if the trigger is set to SINGLE 
then the process freezes after one acquisition that satisfies the RISING or FALLING requirement. This is useful for example in capturing transient signals. 
When the acquisition is stopped, the SAVE DATA button is also available, and one can save the raw signals collected by the oscilloscope, and the corresponding parameters, 
such that the waveforms can be later analyzed with a different software if necessary. 

\section{Testing the oscilloscope}

In order to test the oscilloscope we can use the built in WFG. Thus we just have to connect the output of the WFG channels to the oscilloscope input channels, 
and select the signal type and the parameters. 

In Figure 5 we have a 150KHz sine signal on CHAN0 (WFG0), while on CHAN1 (WFG1) we have a 100KHz triangular signal. 
The zoom is set to 4 for the first signal, and 3 for the second signal, and in both cases the acquisition is performed at $2.4$MS/s.
The cursors A and B are positioned to measure the signal period. For the first signal we have $\Delta \text{BA =0.0067ms, 0.0V}$, while for the second signal we have $\Delta \text{BA =0.01ms, 0.0V}$, 
which are the correct values. In the FFT windows, the peaks are also correctly identified. 

In Figure 6 we have a 50 KHz ramp signal on CHAN0 (WFG0), while on CHAN1 (WFG1) we have a 10KHz square signal. 
For the first signal we have $\Delta \text{BA =0.02ms, 0.0V}$, while for the second signal we have $\Delta \text{BA =0.1ms, 0.0V}$, 
which are the correct values. The FFT peaks are also correctly identified with an error of less than $0.1\%$, including the higher harmonics present at multiples of the fundamental frequency. 

In Figure 7 we have a 75 KHz pulse signal generated by WFG0 and measured on CHAN0, while on CHAN1 we have a random pattern signal generated on WFG1. 
For the pules signal we have the period $\Delta \text{BA =0.01333ms}$, which is the correct value. 
Also, in the FFT window the peak is correctly identified for the pulse signal. The random signal shows a wide band in the Fourier spectrum, as expected.

In a last example we illustrate the PWM functionality by generating a PWM signal with a frequency of 10KHz and a duty of $25\%$ on channel 0, 
and a second PWM signal with 10KHz and a duty of $75\%$ on channel 1. Both PWM signals are captured with a sampling rate of 2,400 KS/s. 
The results are shown in Figure 8. We can see that the frequency is correctly identified in both cases with an error of less than $0.1\%$. 
Also, the FFT power spectrum is identical as it should be, and because TRIGGER1 is set at the same level as TRIGGER0 (1.65 V), 
as RISING on CHAN0, we can see that the signals are actually in perfect synchronization.

The oscilloscope performs quite well for signals with a frequency up to 200KHz. After 200 KHzthe number of measured points is not enough for a proper representation of the signals, and only 
the sine signals can be used (with an acceptable distortion) up to 250KHz. 
Also, the WFG implemented here can generate only positive signals with an amplitude of $0-3.3$V. However the signal acquisition works both in the $0-3.3$V range for DC signals, 
and respectively from $-1.65$V to $+1.65$V for AC signals. Due to these limitations we restricted the frequency settings to a range from 18Hz to 250KHz in the user interface.
Also, because of the large number of visual components embedded on the computer screen, the web software interface requires a minimum screen resolution of 1920x1080 pixels, 
which currently is the standard resolution for most laptop computers today. 

\begin{figure}[H]
\centering \includegraphics[width=6.1in]{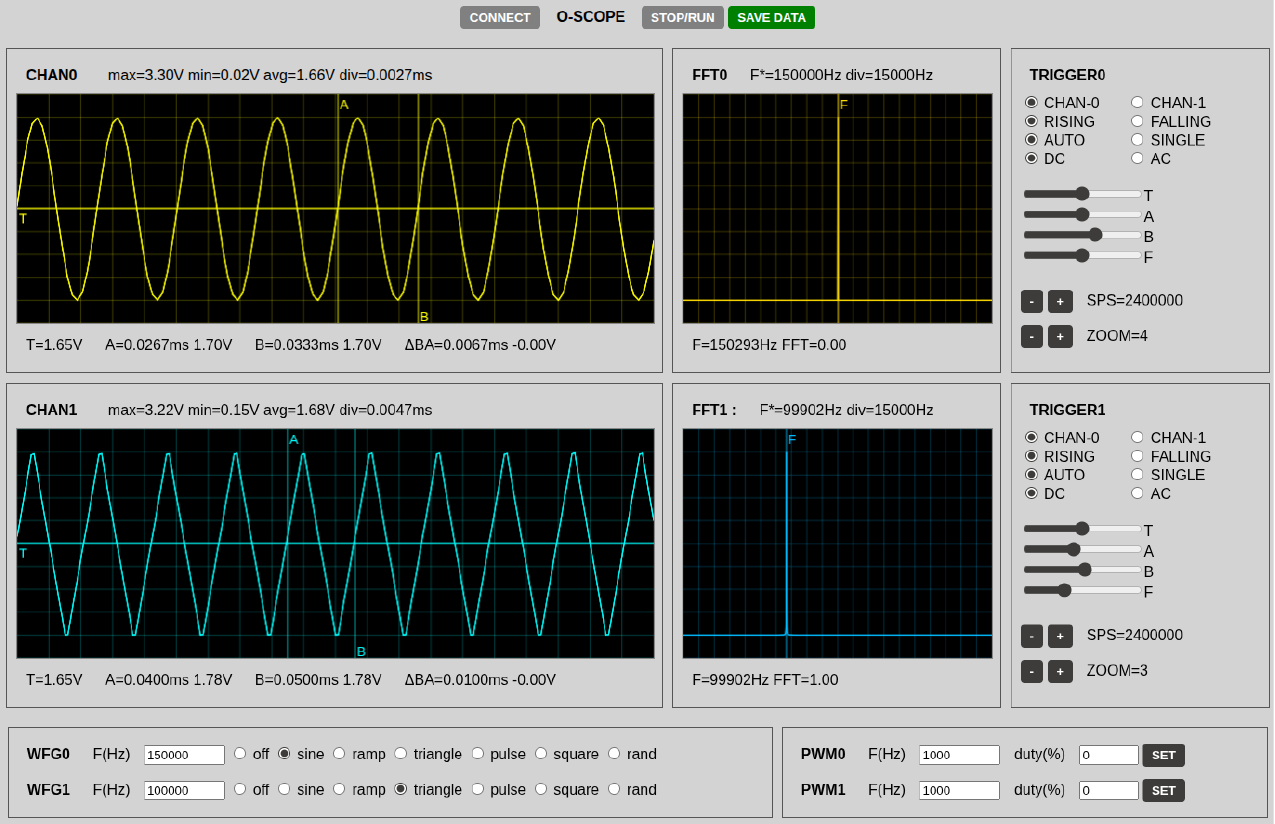}
\caption{CHAN0 (WFG0) sine signal at 150 KHz; CHAN1 (WFG1) triangular signal at 100 KHz.}
\end{figure}

\begin{figure}[H]
\centering \includegraphics[width=6.1in]{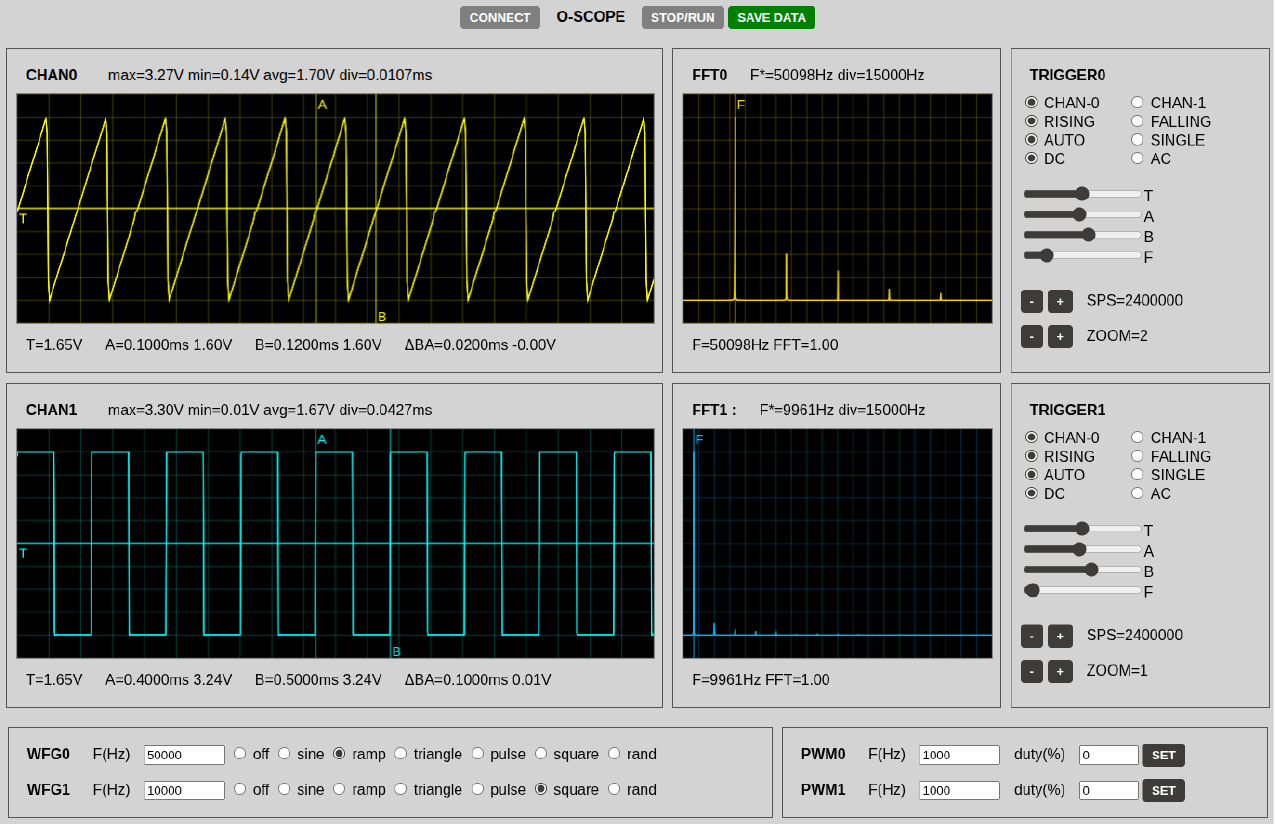}
\caption{CHAN0 (WFG0) ramp signal at 50 KHz; CHAN1 (WFG1) square signal at 10 KHz.}
\end{figure}

\begin{figure}[H]
\centering \includegraphics[width=6.1in]{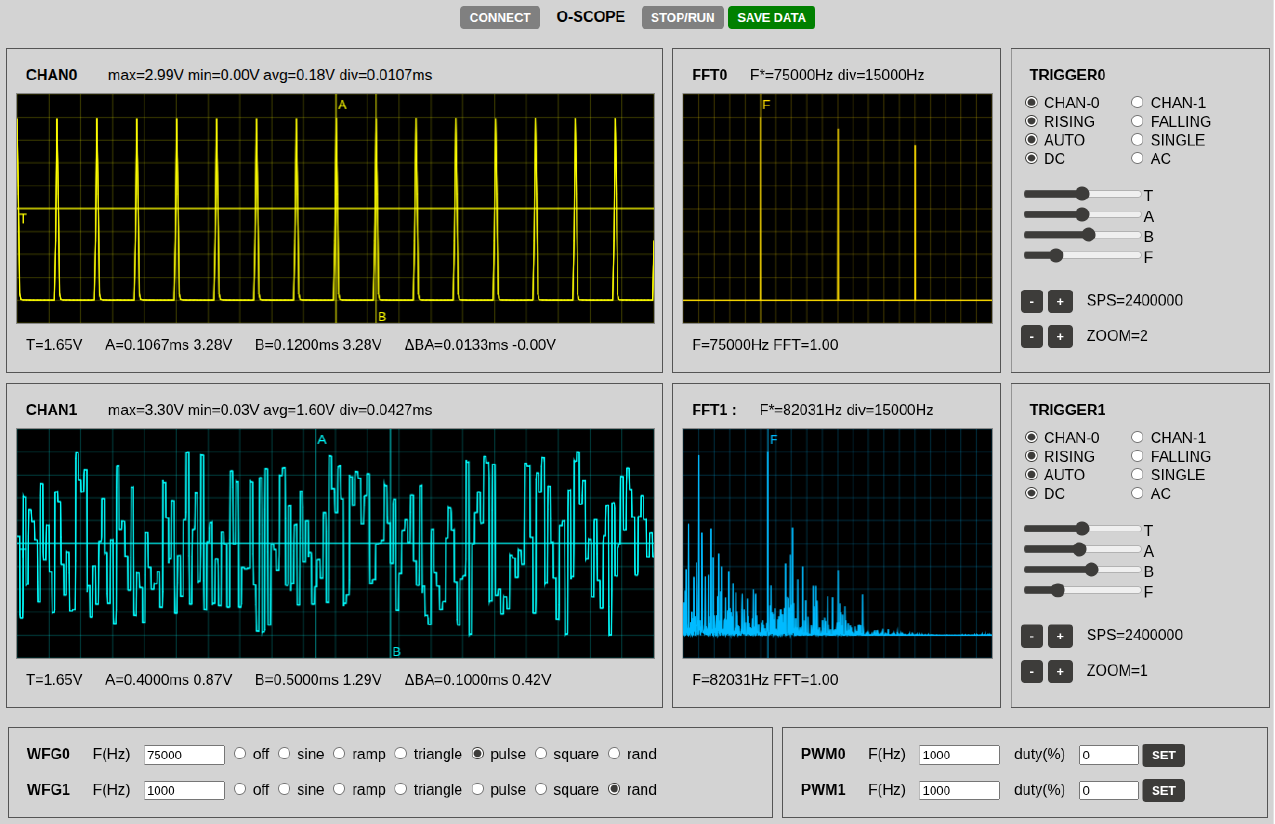}
\caption{CHAN0 (WFG0) Gaussian signal at 75 KHz; CHAN1 (WFG1) random pattern signal.}
\end{figure}

\begin{figure}[H]
\centering \includegraphics[width=6.1in]{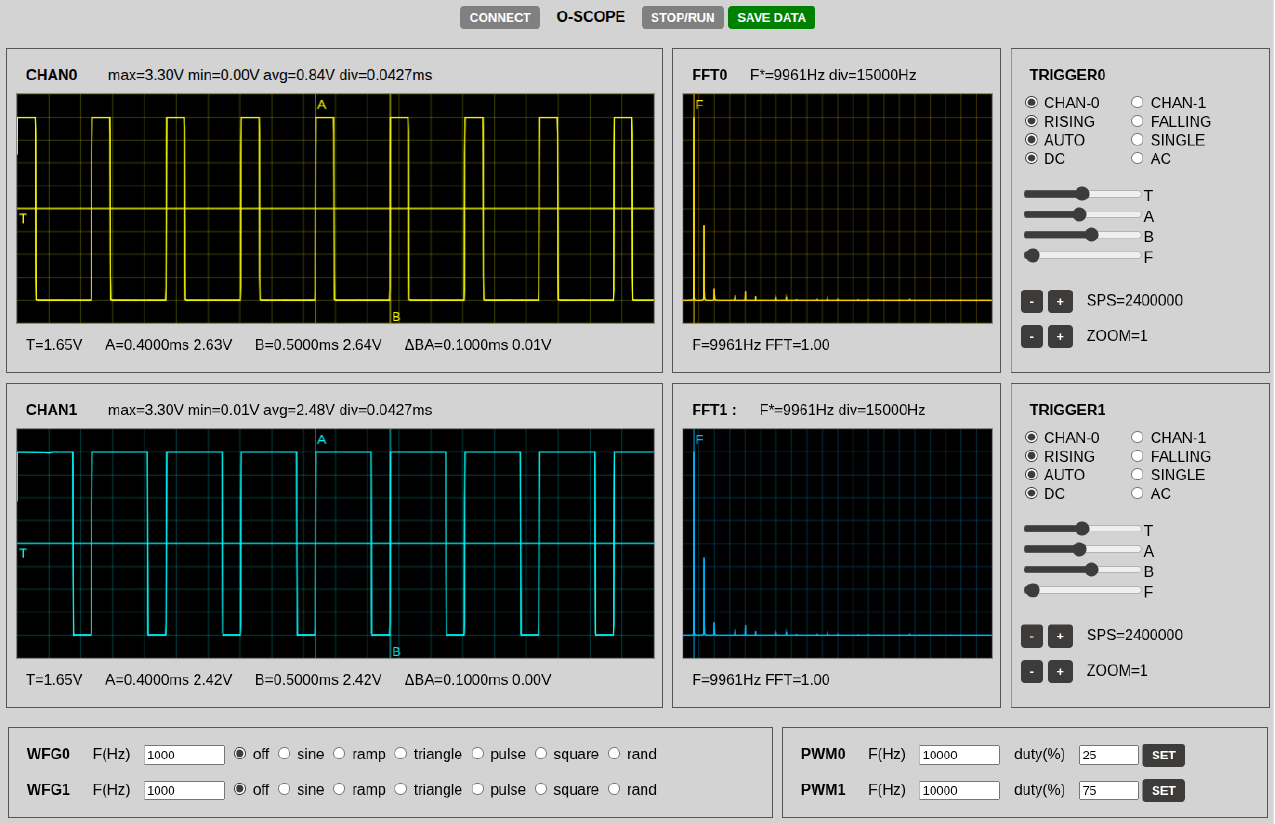}
\caption{CHAN0 PWM0 signal at 10 KHz and duty $25\%$; CHAN1 PWM1 signal at 10 KHz and duty $75\%$;}
\end{figure}

\section*{Conclusion}

We have shown how to build a two channel oscilloscope and a two channel wave function generator (the O-SCOPE), using the RP2 microcontroller boards (RP20350 and RP2040). 
In our implementation we have exploited the PIO and DMA programming in order to achieve high performance digital to analog and analog to digital signal conversions. 
The frontend of the O-SCOPE was built using fast OP-AMP circuits, and it provides a minimal but useful interface between the RP2 boards and the host computer, 
and also it provides input voltage and current protection. 
The software controlling the interface was built on the host computer using vanilla Javascript and the Web Serial API and it runs on the browser. 
The web interface is hosted on the GitHub Pages, and is widely accessible such that no local installations are required. 
The RP2 software component is also freely distributed on GitHub using a UF2 (USB Flashing Format) mass storage file, which can be installed by simply dragging the file on the RP2 boards. 
These features make the O-SCOPE easily accessible, transforming it into a very low cost and useful tool for laboratories everywhere in the world. 

The software is accessible at the following addresses:
\begin{itemize}

\item Web user interface: \url{https://mandrecut.github.io/o_scope/}.

\item O-SCOPE firmware download: \url{https://github.com/mandrecut/o_scope}.

\end{itemize}

The \verb|firmware.uf2| file can be flashed on the RP2 board as following: 

\begin{itemize}

\item Disconnect the RP2 from the micro-USB.

\item Press and hold the BOOTSEL button.

\item Plug the board into the USB port of the host computer and release the button.

\item The computer will mount the RP2 as a mass storage drive named RP2350 (or RP2040).

\item Drag and drop the \verb|firmware.uf2| file onto that drive. 

\item The RP2 board will reboot automatically with the firmware active each time is plugged into an USB port.

\end{itemize}

\end{document}